\documentclass[a4paper,aps,twocolumn,floatfix,showpacs,superscriptaddress]{revtex4-1}

\usepackage{times}
\usepackage{verbatim}
\usepackage{bbold}
\usepackage{bbm}
\usepackage[pdftex]{graphicx}
\usepackage{latexsym,amsmath,verbatim}
\usepackage{xcolor}
\usepackage{hyperref}
\usepackage{lipsum}
\usepackage{rotating}
\usepackage{multirow}
\usepackage[english]{babel}
\usepackage{microtype}
\usepackage{comment}
\usepackage[normalem]{ulem}

\newcommand{\av}[1]{\left\langle {#1} \right\rangle}
\newcommand{\km}{q_\mathrm{max}}

\newcommand{\AV}{\av{q^2}/\av{q}}

\graphicspath{{./}}

\begin{document}

\title{Relevance of backtracking paths in epidemic spreading on networks}

\author{Claudio Castellano}
\email{Corresponding author: claudio.castellano@roma1.infn.it}

\affiliation{Istituto dei Sistemi Complessi (ISC-CNR), Via dei Taurini 
19, I-00185 Roma, Italy}

\author{Romualdo Pastor-Satorras}

\affiliation{Departament de F\'{\i}sica, Universitat Polit\`ecnica de
Catalunya, Campus Nord B4, 08034 Barcelona, Spain}

\begin{abstract} 
  The understanding of epidemics on networks has greatly benefited from
  the recent application of message-passing approaches, which allow to
  derive exact results for irreversible spreading (i.e.  diseases with
  permanent acquired immunity) in locally-tree like topologies.  This
  success has suggested the application of the same approach to
  reversible epidemics, for which an individual can contract the
  epidemic and recover repeatedly.  The underlying assumption is that
  backtracking paths (i.e. an individual is reinfected by a neighbor
  he/she previously infected) do not play a relevant role.  In this
  paper we show that this is not the case for reversible epidemics,
  since the neglect of backtracking paths leads to a formula for the
  epidemic threshold that is qualitatively incorrect in the large size
  limit.  Moreover we define a modified reversible dynamics which
  explicitly forbids direct backtracking events and show that this
  modification completely upsets the phenomenology.
\end{abstract}

\maketitle

\section{Introduction}
\label{sec:introduction}

The study of epidemic processes~\cite{Pastor-Satorras:2014aa} has been
one of the fundamental levers in the development of modern network
science~\cite{Newman10}. Research in this field has been based on the
consideration of several simplified epidemic models~\cite{anderson92},
the most fundamental being the susceptible-infected-susceptible (SIS)
and the susceptible-infected-removed (SIR) models. In both of them, a
susceptible (healthy) individual, in contact through an edge with an
infected individual, contracts the disease at rate $\beta$. In the SIS
model infected nodes become spontaneously susceptible again at rate
$\mu$, while in the SIR model infected individuals become removed, and
cannot contract the disease again.  This variation results in a
different collective behavior: SIS allows for a long lived (infinitely
long in the thermodynamic limit of infinite network size) steady state
with a finite fraction of infected individuals, while the SIR model
proceeds by epidemic outbreaks of finite duration, that reach a given
fraction of individuals in the population. Both these models are
characterized by the presence of an epidemic threshold $\lambda_c$ for
the control parameter, defined as the ratio between the infection and
the healing rates, $\lambda = \beta / \mu$. In the SIS model, the region
$\lambda \leq \lambda_c$ corresponds to the healthy phase, in which
epidemic episodes die out exponentially fast; for $\lambda > \lambda_c$,
on the other hand, a steady state with a finite fraction of infected
individuals is observed. For the SIR model, on the other hand, the
healthy phase corresponds to outbreaks affecting an infinitesimally
small fraction of the population, while for $\lambda > \lambda_c$ the
final fraction of removed individuals becomes finite.

Initial analytical understanding of epidemic processes on networks was
based on the so-called heterogeneous mean-field (HMF)
approach~\cite{pv01a}, which assumes that the network is
annealed~\cite{dorogovtsev07:_critic_phenom} and topologically
uncorrelated~\cite{assortative}.  Under these assumptions, both the SIR
and the SIS process exhibit a threshold $\lambda^\mathrm{HMF}_c$
inversely proportional to the second moment $\av{q^2}$ of the degree
distribution $P(q)$ characterizing the topology of the
network~\cite{Newman10}. For a distribution scaling as a power-law for
large degree, $P(q) \sim q^{-\gamma}$, as empirically observed in many
contexts~\cite{barabasi02}, this second moment scales, for
$\gamma \leq 3$, as $\av{q^2} \sim \km^{3-\gamma}$, where $\km$ is the
maximum degree in the network~\cite{Dorogovtsev:2002}. In this regime
$\gamma \leq 3$, the maximum degree diverges in the limit of infinite
network size, leading to a vanishingly small epidemic threshold. For
$\gamma >3$, instead, both the second moment and the threshold attain a
finite value in the thermodynamic limit.

For the SIS model, a more refined approach, the quenched mean-field
(QMF) theory~\cite{Wang03,Castellano2010,PVM_ToN_VirusSpread,Gomez10},
predicts an epidemic threshold $\lambda^\mathrm{QMF}_c = 1/\Lambda_M^A$,
where $\Lambda_M^A$ is the largest eigenvalue (LEV) of the adjacency
matrix $A_{ij}$ representing the structure of connections in the
network~\footnote{The symmetric adjacency matrix takes values
  $A_{ij} = 1$ if nodes $i$ and $j$ are connected by an undirected edge,
  and $A_{ij} = 0$ otherwise.}. The observation that, in general
networks with a power-law degree distribution, $\Lambda_M^A$ diverges
with network size~\cite{Chung03,Castellano2017}, leads to a vanishing
threshold for any value of the degree exponent $\gamma$, in
contradiction with HMF theory.  The validity of QMF theory has been
numerically checked~\cite{Ferreira2012} and later extended, by taking
into account long distance reinfection events between high degree
nodes~\cite{PhysRevLett.111.068701}.

For the case of the SIR model, a new theoretical framework has been
recently proposed, based on the message passing (MP)
approach~\cite{Karrer2010}. This approach builds on the observation that
a node $i$, that has been infected by a neighbor node $j$, has
absolutely no chance to reinfect $j$.  This amounts to disregarding, in
the epidemic evolution, the possibility of \textit{backtracking} paths
allowing for the mutual reinfection of pairs of connected nodes.  A
theory based on the mapping of the SIR model onto bond
percolation~\cite{Grassberger1983} leads to an epidemic threshold for
the SIR model given by the inverse of the largest eigenvalue
$\Lambda_M^H$ of the Hashimoto or non-backtracking
matrix~\cite{Karrer2014,Hashimoto1989211}.  This matrix has also been
applied to the definition of centrality measures in
networks~\cite{2014arXiv1401.5093M} and to the detection of
communities~\cite{Krzakala2013}.  The validity of the MP method for
percolation and SIR dynamics has been numerically
confirmed~\cite{PhysRevE.91.010801}.

In a recent paper, Shrestha \textit{et al.}~\cite{Shrestha2015}, apply
the same MP approach to the SIS model, arguing that backtracking paths
lead to ``echo chamber'' effects (successive reinfection events between
neighboring nodes) and those effects should be disregarded as somehow
pathological.  Application of the MP approach to the SIS model provides
a set of differential equations that can be solved to obtain the
evolution of the prevalence (density of infected individuals) as a
function of time, and leads to an epidemic threshold again inversely
proportional to the largest eigenvalue $\Lambda_M^H$ of the
non-backtracking matrix.  Based on numerical simulations of the SIS
process on a variety of small real and synthetic networks, the authors
of Ref.~\cite{Shrestha2015} conclude that their MP approach provides an
accurate estimate of the evolution of prevalence with time, as it takes
properly into account dynamical correlations (neglected in both HMF and
QMF approaches), as well as a better bound for the epidemic threshold.

The results of Ref.~\cite{Shrestha2015} are in striking contradiction
with previously published literature and the physical picture emerging
from it. In particular, Ref.~\cite{Castellano2012} pointed out the
existence of different mechanisms triggering the onset of the extended
and long-lasting outbreaks, highlighting the crucial role played in some
cases by the subset composed by the node with the largest degree (the
hub) and its immediate neighbors.  This star graph alone is sometimes
able to self-sustain the epidemic and spread it to the rest of the
system: Backtracking events, with the repeated mutual reinfection among
the hub and the leaves of the star graph, are at the heart of this
phenomenology.  The physical picture introduced in
Ref.~\cite{Castellano2012} has been confirmed
elsewhere~\cite{Ferreira2012,PastorSatorras2018}.  A clarification of
the apparent contradiction with respect to Ref.~\cite{Shrestha2015} is
therefore needed.

In this paper we show that, even though in some cases a MP approach
might provide a good approximation to the value of the prevalence, it
does not account properly for the position of the epidemic threshold in
power-law distributed networks, specially in the case of very large
networks. Comparing the predictions of QMF and MP theories, as given by
the inverse of the largest eigenvalues of the adjacency and Hashimoto
matrices, and comparing them with direct large scale numerical
simulations of the SIS model on real and uncorrelated synthetic
networks, we show that MP theory does not capture the correct behavior
of the epidemic threshold, particularly for large values of the degree
exponent $\gamma$.  We also observe that the MP prediction for the
threshold is not a bound for the true value, while the scaling of the
threshold with network size is instead more accurately described by QMF
theory. Moreover we consider a modified SIS model in which backtracking
is hindered, and show that its behavior radically differs from the
original SIS model.  Our work shows that backtracking paths, at odds
with the claims in Ref.~\cite{Shrestha2015}, do play a fundamental role
in the dynamics of the SIS model, and are indeed at the core of the
vanishing threshold observed asymptotically for uncorrelated networks
with power-law distributed degree.

\section{Comparison of QMF and MP predictions with numerical
  simulations} 
\label{sec:direct-comp-betw}

The Quenched Mean-Field theory predicts the SIS threshold to be
inversely proportional to the largest eigenvalue $\Lambda_M^A$ of the
adjacency matrix.  The Message Passing approach gives an analogous
formula with the only difference that the matrix for which the largest
eigenvalue must be calculated is the Hashimoto matrix. In order to
visualize the comparison between both predictions and also see what
happens when $N$ is not large, we evaluate numerically the largest
eigenvalue of the adjacency and Hashimoto matrices.  For the adjacency
matrix, $\Lambda_M^A$ is calculated by applying a simple power iteration
strategy~\cite{golub2012matrix}. In the case of the Hashimoto matrix,
$\Lambda_M^H$ is evaluated by using the Ihara-Bass determinant
formula~\cite{2014arXiv1401.5093M}, i.e., by computing the largest
eigenvalue of the $2N \times 2N$ matrix
\begin{equation}
  \label{eq:1}
  \mathbf{M} = \left(
    \begin{matrix}
      \mathbf{A} & \mathbf{I} - \mathbf{D}\\
      \mathbf{I} & \mathbf{0}
    \end{matrix}
\right),
\end{equation}
where $\mathbf{A}$ is the adjacency matrix, $\mathbf{I}$ is the identity
matrix, $\mathbf{D}$ is a diagonal matrix with $D_{ii} = q_i$, the
degree of node $i$, and $\mathbf{0}$ is the null matrix. The largest
eigenvalue of the matrix $\mathbf{M}$ is again computed using the power
iteration method.

To compare the theoretical predictions with actual values of the SIS
threshold, we compute the latter using the lifespan
method~\cite{PhysRevLett.111.068701,Mata15}.  In this approach,
simulations start with only the hub infected.  For each run, one keeps
track of the coverage $c$, defined as the fraction of different nodes
that have been infected at least once. In an infinite network, this
quantity is vanishing for $\lambda < \lambda_c$, while it tends
asymptotically to $1$ in the active region of the phase diagram. In
finite networks, one can set a coverage threshold $c_t$ and consider all
runs that reach $c \geq c_t$ as endemic.  The average lifespan $\av{T}$,
where averaging is restricted only to nonendemic runs, plays the role of
a susceptibility: The position of the threshold is estimated as the
value of $\lambda$ for which $\av{T}$ reaches a peak. In our simulations
we choose $c_t = 1/2$.

\subsection{Real networks}
\label{sec:real-networks}

We start our analysis considering the set of $109$ real networks
described in Ref.~\cite{Radicchi15}. This set represents a collection of
networks of widely varying size, heterogeneity and level of topological
correlations~\cite{alexei,assortative}, see Ref.~\cite{Castellano2017}.
\begin{figure}[t]
  \centering
  \includegraphics[width=\columnwidth]{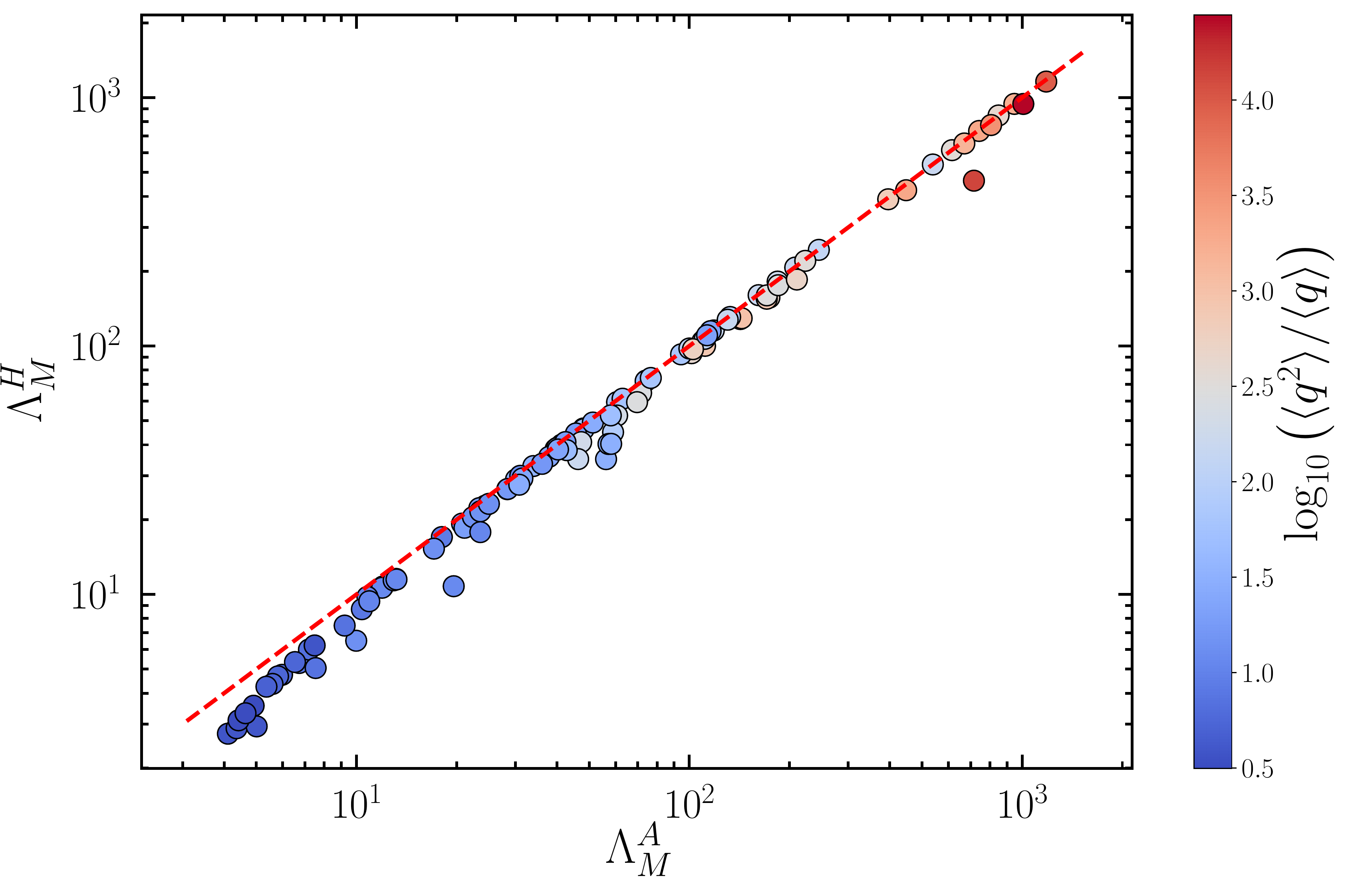}
  \caption{Scatter plot of the Hashimoto matrix LEV $\Lambda_M^H$ vs the
    LEV of the adjacency matrix $\Lambda_M^A$ for the $109$ real
    networks described in Ref.~\cite{Radicchi15}. The color of the
    symbols encodes the network heterogeneity as measured by the factor
    $\av{q^2}/\av{q}$.}
  \label{fig:scatterLEVsrealnets}
\end{figure}
In Fig.~\ref{fig:scatterLEVsrealnets} we present a comparison of the
numerically estimated values of the LEVs $\Lambda_M^A$ and
$\Lambda_M^H$, presented as a scatter plot. As we can see from this
figure, the difference between the two LEVs is minimal, being almost
equal with the exception of the networks with a low level of
heterogeneity, as measured by the factor $\av{q^2}/\av{q}$. Thus, only
for small values of $\av{q^2}/\av{q}$, the LEV of the Hashimoto matrix
is noticeably smaller than $\Lambda_M^A$.  It is interesting to notice
that this behavior is not clearly correlated with network size (data not
shown), since some homogeneous networks with large size show marked
differences in the their LEVs.

\begin{figure}[t]
  \centering
  \includegraphics[width=\columnwidth]{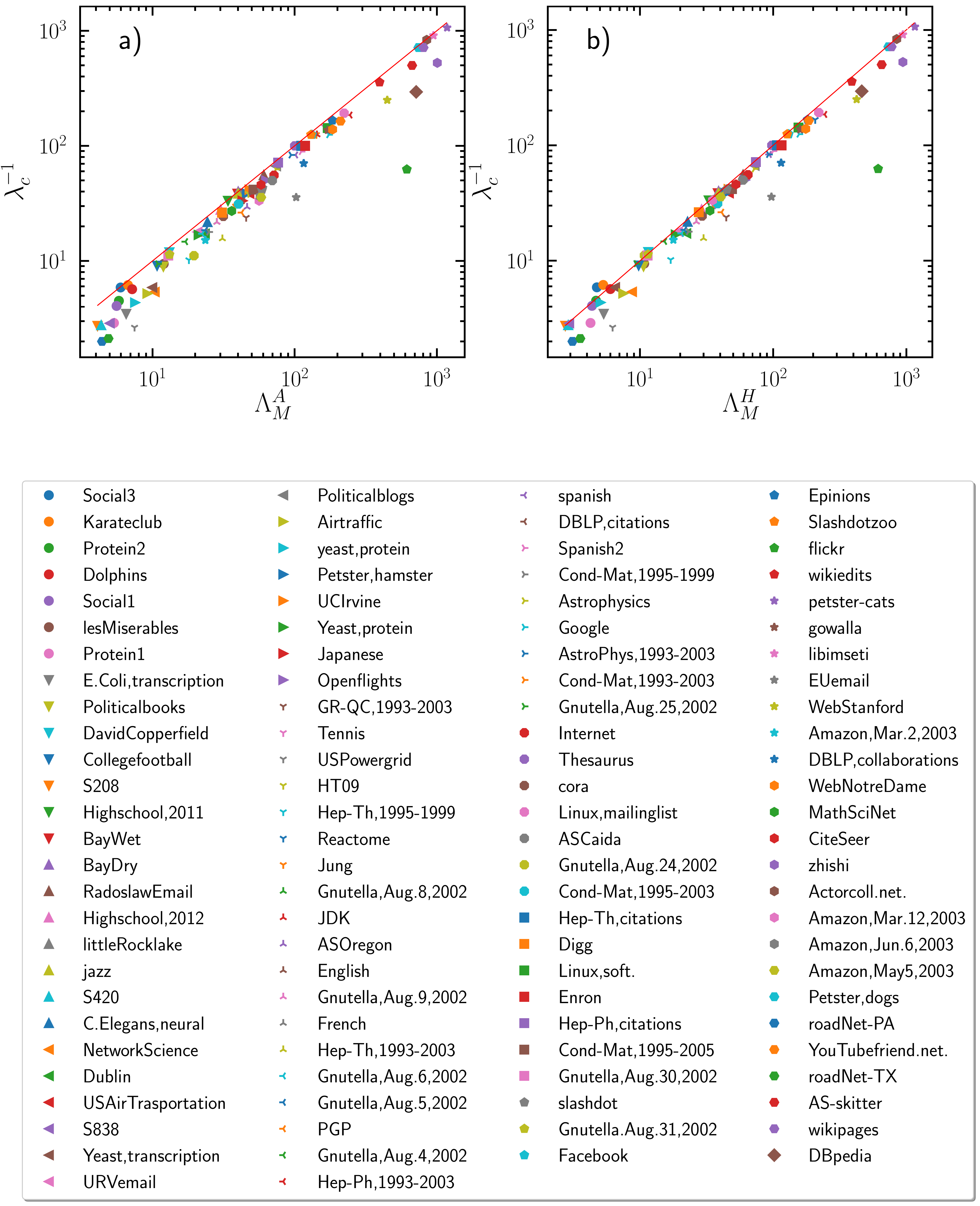}
  \caption{Scatter plot of the inverse numerical SIS threshold estimated
    for the real networks considered, as a function of the
    QMF prediction $\Lambda_M^A$ (a) and  the MP prediction
    $\Lambda_M^H$ (b). Solid lines represent perfect agreement.}
  \label{fig:SIScomparisonrealnets}
\end{figure}

Because of the similarity among the two predictions it is difficult to
assess which of the two better reproduces numerical SIS thresholds.  We
present such a comparison in Fig.~\ref{fig:SIScomparisonrealnets}, where
we plot the numerical SIS threshold $\lambda_c$, computed for all
networks of the dataset considered (except the three largest), as a
function of the QMF and MP predictions.
As we can see from the Figure, the relative accuracy of the two theoretical
predictions is practically the same.  The MP theory seems to be marginally
better in some cases, but this is a consequence of the fact that the MP
prediction (contrary to the QMF prediction) is not a strict bound on the
actual threshold.

\subsection{Synthetic uncorrelated networks}
\label{sec:synthetic-networks}

While the consideration of real networks is undoubtedly interesting, it
suffers from the problem that real networks are topologically complex,
being rife with correlations, clustering and non-trivial community
structures~\cite{Newman10}, which are not taken into account in
theoretical approaches. For this reason, we now turn to the analysis of
synthetic uncorrelated networks, lacking all those topological
complications. In the case of uncorrelated networks, Chung and
collaborators~\cite{Chung03} provide an expression for the LEV of the
adjacency matrix that can be safely
recast~\cite{Castellano2017,Goltsev2012} as
\begin{equation}
\Lambda_M^A \approx \max \{ \sqrt{\km}, \av{q^2}/\av{q}\}.
\label{CLV}
\end{equation}
For the Hashimoto matrix instead, the value of the LEV can be
approximated
by~\cite{Krzakala2013,2014arXiv1401.5093M,PhysRevE.91.010801}
\begin{equation}
\Lambda_M^H \approx \AV-1 .
\label{Krzakala}
\end{equation}

It is immediate to notice that the two expressions are practically the
same if the max in Eq.~(\ref{CLV}) is given by the second argument and
$\av{q^2}/\av{q}$ is sufficiently large. For power-law distributed
networks of size $N$ this always occurs in the large $N$ limit, provided
$\gamma<5/2$.  On the other hand it is clear that the two expressions
give a qualitatively different result when $\sqrt{\km}$ is largest in
Eq.~(\ref{CLV}), i.e. for $\gamma>5/2$ and large
$N$~\cite{Castellano2010,Goltsev2012}.

\begin{figure}[t]
  \centering
  \includegraphics[width=\columnwidth]{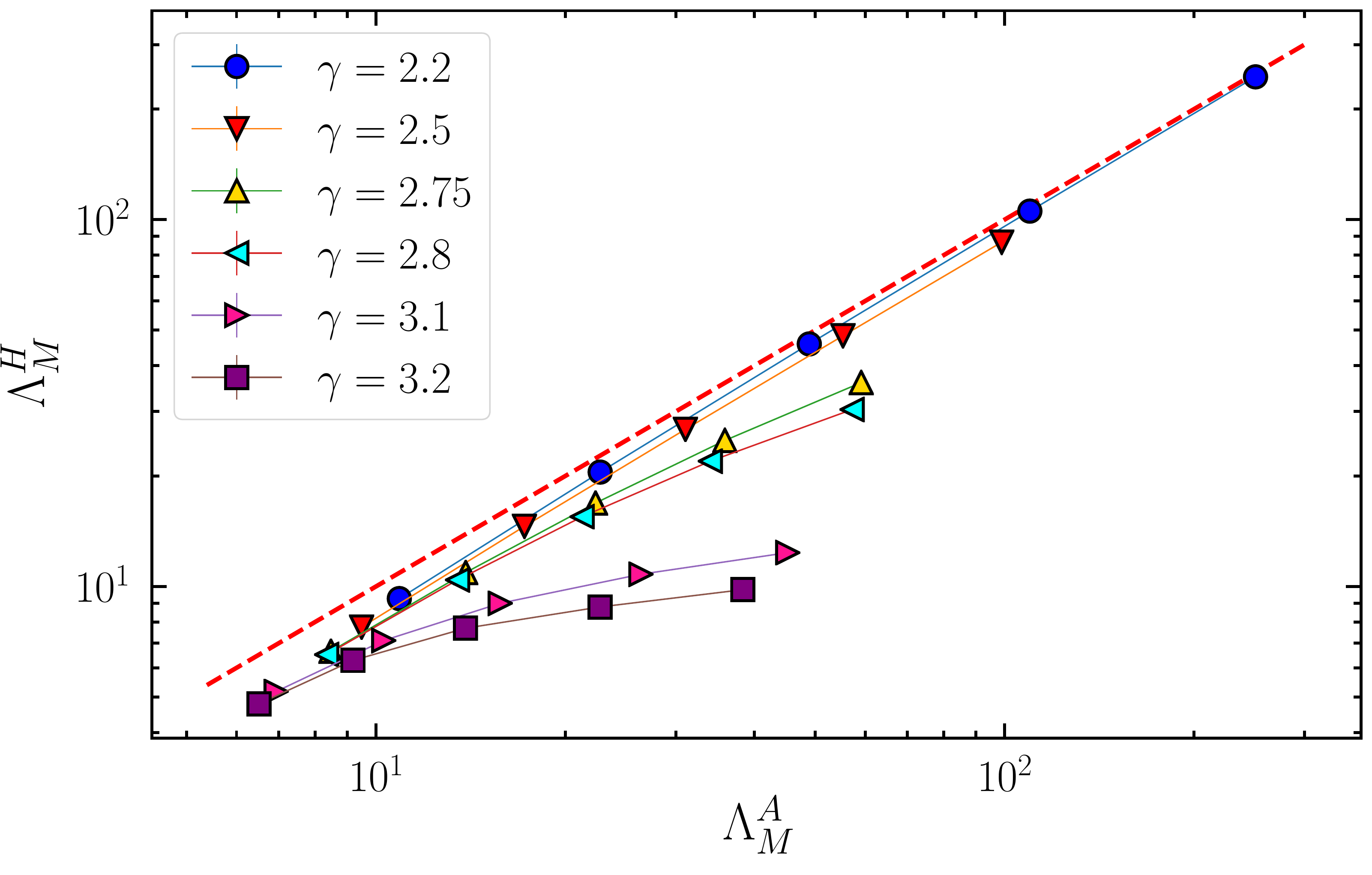}
  \caption{Scatter plot of the Hashimoto matrix LEV $\Lambda_M^H$ vs the
    LEV of the adjacency matrix $\Lambda_M^A$ for power-law UCM networks
    of varying size from $N=10^3$ to $N=10^7$, averaged over $100$
    different network realizations for each value of $N$. Error bars are
    smaller than symbol sizes.}
  \label{fig:scatterLEVs}
\end{figure}
To investigate what happens also for moderate network size, we numerically
compute the eigenvalues $\Lambda_M^A$ and $\Lambda_M^H$ for uncorrelated
synthetic power-law networks generated using the uncorrelated configuration
model (UCM)~\cite{Catanzaro05}, with minimum degree $q_\mathrm{min} = 3$, and
a maximum degree $q_\mathrm{max} = N^{1/2}$ for $\gamma \leq 3$ (to avoid
degree correlations~\cite{mariancutofss}) and $q_\mathrm{max} =
N^{1/(\gamma-1)}$ for $\gamma > 3$ (to avoid large sample-to-sample
fluctuations~\cite{Boguna09}). In Fig.~\ref{fig:scatterLEVs} we present a
scatter plot of the largest eigenvalue of the Hashimoto matrix as a function
of $\Lambda_M^A$ for various degree exponents $\gamma$ and network sizes
ranging from $N=10^3$ to $10^7$.

As expected, for $\gamma \leq 5/2$ the largest eigenvalues of both matrices
attain essentially the same value in the limit of large network size; in this
case, the predictions of QMF and MP theories are practically
indistinguishable. For $\gamma>5/2$, on the other hand, we observe that
$\Lambda_M^H$ becomes much smaller than $\Lambda_M^A$ as larger values of $N$
are considered. From Eqs.~(\ref{CLV}) and~(\ref{Krzakala}), we know that
asymptotically the former reaches a constant value, while the latter diverges.
This represents a strong contradiction between the two theories: while QMF
predicts a vanishing threshold for all $\gamma$, MP gives a finite threshold
for any $\gamma > 5/2$.  Notice also that for small values of $N$ the two
eigenvalues are not  equal, but also not very different.

This analysis of synthetic networks is in agreement with our observations on
real ones. Highly heterogeneous networks have large and practically identical
LEVs. On the other hand, for networks with low heterogeneity, the LEVs are
dissimilar, with $\Lambda_M^H$ in general smaller than $\Lambda_M^A$. 

\begin{figure}[t]
  \centering
  \includegraphics[width=\columnwidth]{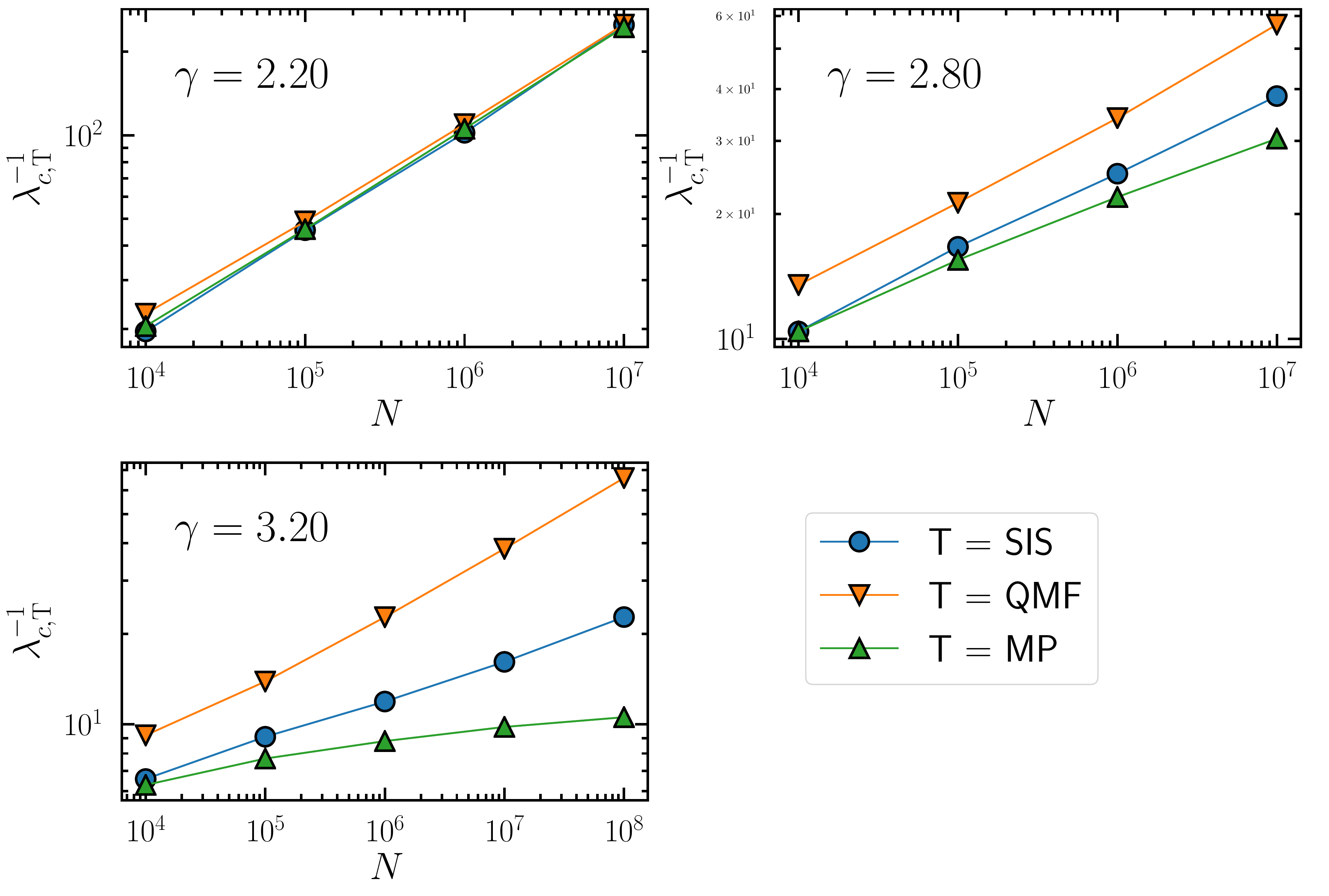}
  \caption{Comparison of the scaling of the inverse numerical SIS
    threshold, $\lambda_{c, \mathrm{SIS}}$, the QMF prediction,
    $\lambda_{c, \mathrm{QMF}}$, and the MP prediction
    $\lambda_{c, \mathrm{MP}}$ as a function of network size for
    different values of the degree exponent $\gamma$.}
      \label{fig:SISsimulations}
\end{figure}

In order to compare the validity of the two theoretical approaches with
respect to the epidemic threshold of the SIS model, in
Fig.~\ref{fig:SISsimulations} we plot the inverse numerical epidemic
threshold, estimated from the lifespan method, and the inverse of the QMF and
MP predictions as a function of network size, on uncorrelated networks
generated with the UCM algorithm for different values of $\gamma$.
From this figure it turns immediately out that for $\gamma=2.2$ the
predictions of the two theories essentially coincide and reproduce extremely
well numerical results. Indeed, the MP prediction seems in this case to work
slightly better than QMF. This is due to the additional term $-1$ in
Eq.~(\ref{Krzakala}) with respect to Eq.~(\ref{CLV}).  This term accounts for
dynamical correlations, that in general prevent a node from infecting the node
that transmitted the disease to it in the first place, as the latter is with
high probability still infected~\cite{PhysRevLett.111.068701}.  The opposite
occurs, on the other hand, for $\gamma > 5/2$. In the case $\gamma=2.8$, both
the inverse numerical SIS threshold and the QMF and MP predictions diverge in
the large size limit, but the slope of the growth of the SIS threshold is
better captured by QMF. For $\gamma=3.2$, the MP prediction tends to a
constant value, while the inverse SIS threshold grows as a power law, in the
same fashion as the inverse QMF prediction.  Notice also that while the QMF
prediction is an upper bound for the inverse threshold, this is clearly not
the case for the MP prediction.

We conclude that while both theories perform extremely well for heterogeneous
networks with small values of $\gamma$, for less heterogeneous networks they
are both inaccurate.  However, QMF captures the fundamental feature that the
threshold vanishes in the large size limit, while the neglect of backtracking
events has the consequence that MP is qualitatively off target.

\section{A SIS-like model without direct backtracking}
\label{sec:sis-model-without}

In order to further clarify the relevance of backtracking paths in the SIS
dynamics, we consider a modified SIS-like dynamics, in which such paths are
strongly depressed.

First of all, let's point out that if backtracking is fully ruled out, no
steady state is ever possible. Indeed, in that case, each infection event
practically ``removes'' the corresponding edge, which is not available any
more for transmitting the infection. In the long run all edges are removed and
the absorbing state with all nodes in state S is the only possible asymptotic
configuration.  Therefore we consider a modified SIS dynamics where
backtracking events are prohibited, but only until other infection events take
place.  Our model, that we dub Non-Backtracking SIS (NBSIS) dynamics, is
defined in terms of the SIS dynamics, with the following addition: if at a
certain time node $j$ is infected by node $i$ and then $i$ becomes
susceptible, $j$ cannot transmit the epidemic back to node $i$ before that
some other neighbor $m$ of $i$ ($m \neq j$) reinfects it.
\begin{figure}[t]
  \centering
  \includegraphics[width=\columnwidth]{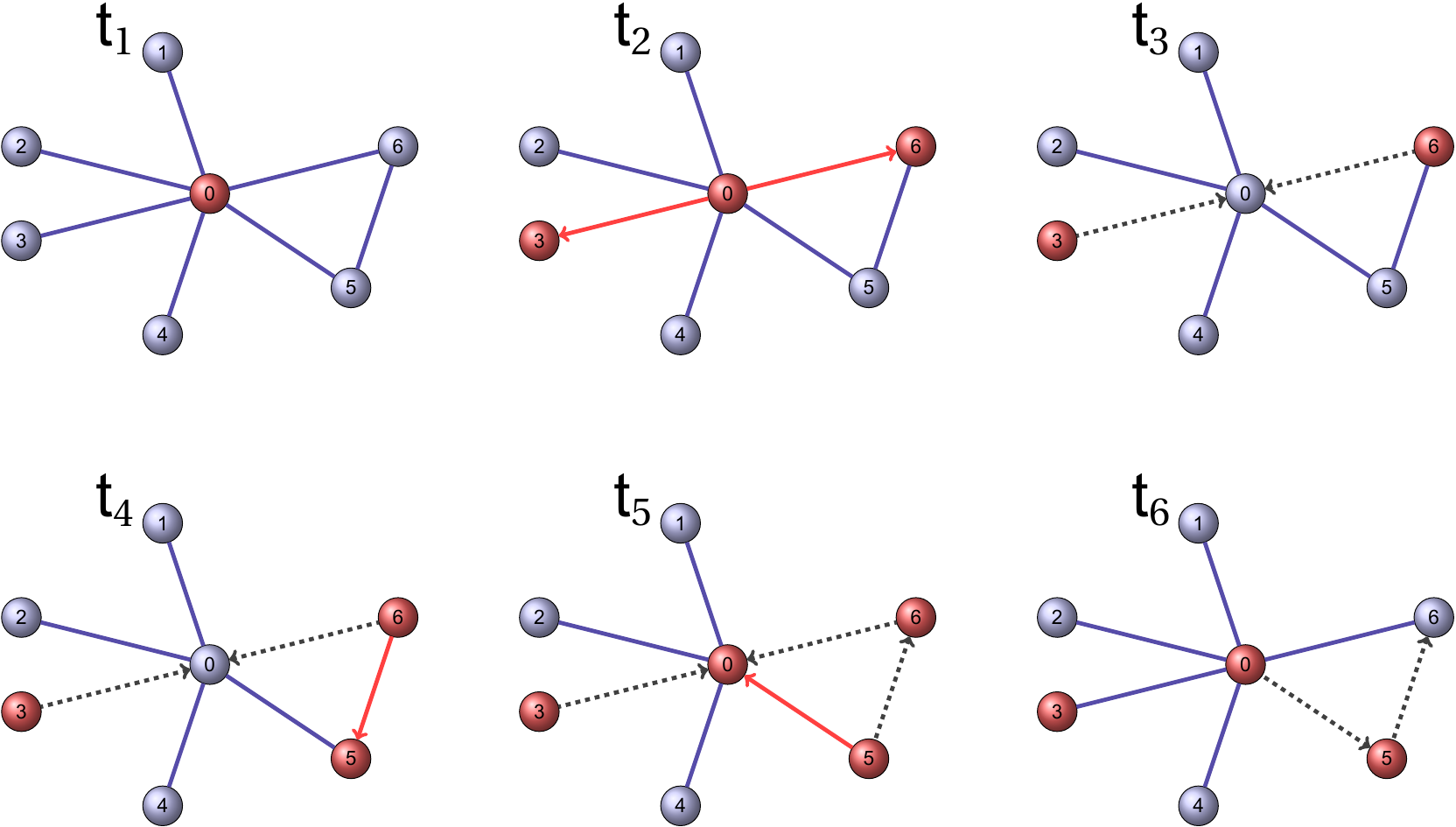}
  \caption{Illustration of allowed and disallowed events in a time
    sequence of the NBSIS dynamics. Red nodes are in state I, blue nodes
    are in state S.  Panels are ordered temporally.  Solid blue edges
    are those allowing the transmission of the infection. Red arrows
    represent infection events. Black dashed arrows do not allow the
    transmission in the corresponding direction.}
  \label{fig:NBSISdefinition}
\end{figure}
An illustration of this type of dynamics is provided in
Fig.~\ref{fig:NBSISdefinition}.  The central node is initially infected; it
transmits the disease to two of its neighbors and then recovers. The two
neighbors cannot retransmit the infection back to it. However they can
transmit the infection to other neighbors of theirs, that can, in their turn,
reinfect the central node.  With this recipe, local ``echo chamber'' effects
are strongly depressed.  This small variation can have important effects even
in simple network structures. In the case of a star network, composed by a hub
of degree $q$ connected to $q$ leaves of degree $1$, it is trivial to observe
that the NBSIS model does not allow a steady state. Indeed, leaves can only be
infected by the hub; therefore, once a leaf has been infected, it cannot be
infected by any other node, an thus after it spontaneously recovers, it has no
chance to be reinfected again. The NBSIS epidemic dies out in a few time steps
after its onset. This behavior is in contrast with the SIS dynamics on star
networks, which can sustain long-lived steady states as soon as $\lambda >
1/\sqrt{q}$ \cite{Ganesh05}. The same lack of an active state is observed in
generic tree networks~\cite{Newman10} when the infection starts at a single
node.

In the case of generic networks, again a lack of steady state is observed in
the limit of large $\lambda$.  Indeed, in the limit $\lambda \to \infty$, the
SIS model quickly reaches a clear steady state in which the density of
infected nodes tends to one. In the NBSIS model, on the other hand, if
$\lambda \to \infty$, the initial infected seed infects all its nearest
neighbors in the initial time steps. These infected neighbors cannot reinfect
the seed, due to the non-backtracking condition, and therefore the seed, once
recovered, cannot become infected again and thus becomes effectively removed
from the network. The neighbors of the seed experience a similar fate, after
infecting all their neighbors. Therefore, in this limit the NBSIS dynamics
lacks an active (infected) phase, and quickly decays to the absorbing
(healthy) phase.

\begin{figure}
  \centering
  \includegraphics[width=\columnwidth]{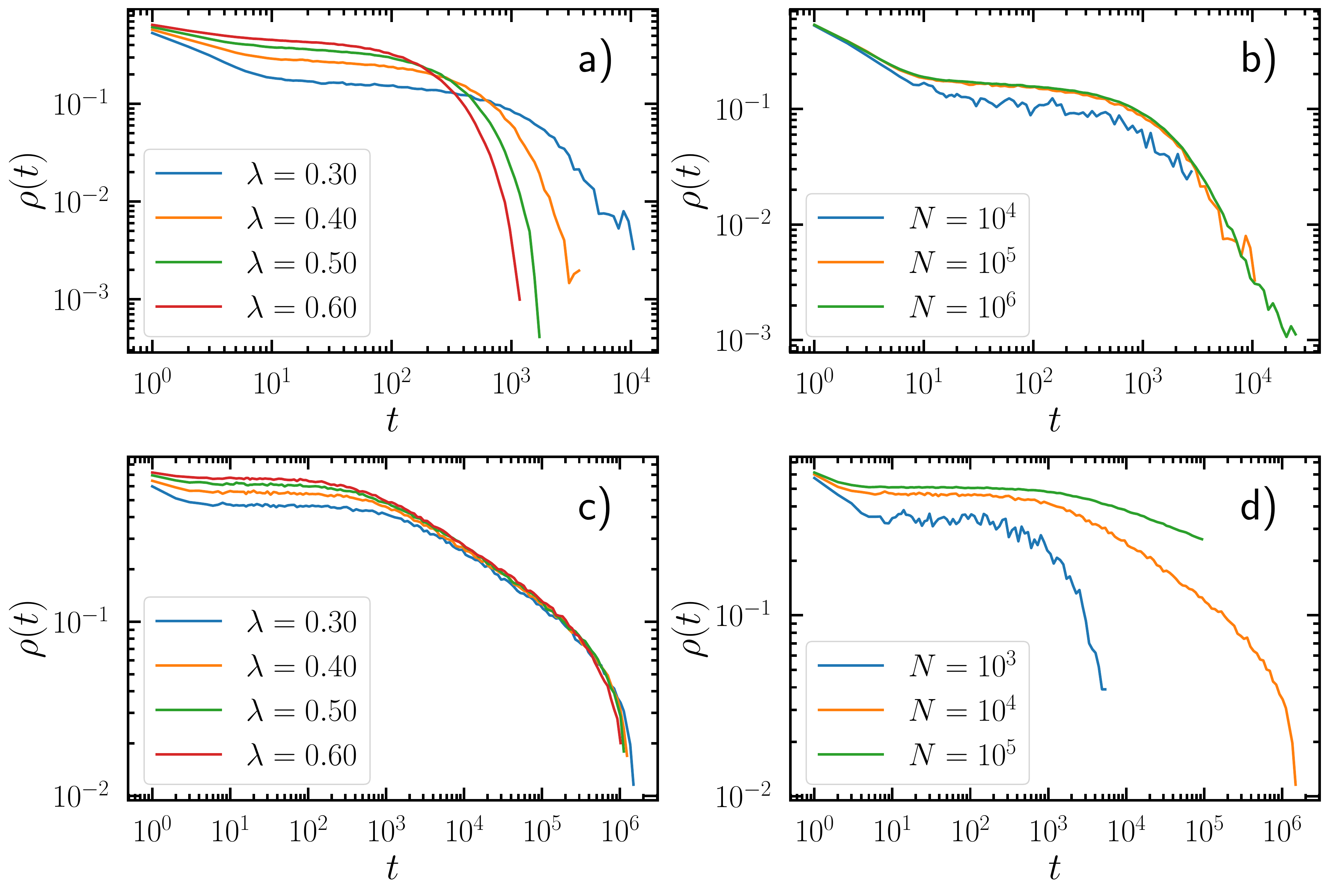}
  \caption{a) Prevalence $\rho(t)$ as a function of time for
    $\gamma=3.2$ and various values of $\lambda$, $N=10^5$. b) Same plot
    for $\lambda=0.3$ and changing network size $N$. c) Prevalence
    $\rho(t)$ as a function of time for $\gamma=2.2$ and various values
    of $\lambda$, $N=10^4$. d) Same plot for $\lambda=0.3$ and changing
    network size $N$. Results presented are obtained for a single
    epidemic run at a given value of $\lambda$ and $N$.}
  \label{Prevalence}
\end{figure}

In Fig.~\ref{Prevalence} we show that the lack of an active state occurs for
any finite value of lambda in the NBSIS, which shows instead a transient that
eventually decays into the absorbing state.
As we can observe in Fig.~\ref{Prevalence}(a), for $\gamma > 5/2$ the
density of infected individuals tends to zero for a sufficiently large
time interval, whose length {\em decreases} as $\lambda$ is increased,
in stark contrast with what is expected for a system with a truly active
state above its critical point.  Moreover, as Fig.~\ref{Prevalence}(b)
shows, this time interval does not depend on the system size, revealing
that for any value of $\lambda$ the only stable state is the absorbing
one. In the case $\gamma < 5/2$ [Fig.~\ref{Prevalence}(c)] we observe
some differences: The initial plateau is followed by an intermediate
regime with a decay slower than exponential. For fixed $\lambda$ and
increasing network size $N$, Fig.~\ref{Prevalence}(d), the decay toward
the absorbing state occurs over temporal scales that grow with $N$.
Still the most important features are the same: There is no steady state
and the temporal scale over which the epidemic disappears (weakly)
decreases with $\lambda$~\cite{PhysRevX.5.031017}.

Overall, this phenomenology can be attributed to the suppression of
backtracking paths for the infection. In all cases no steady state is
sustained for asymptotically long times.  Notice that the way the
absorbing state is reached for finite networks is completely different
from what happens for SIS.  In the latter case the absorbing state is
reached from the steady state by a random fluctuation; for NBSIS instead
$\rho(t)$ is pushed toward zero by a deterministic drift.

\begin{figure}[t]
 \centering

 \includegraphics[width=\columnwidth]{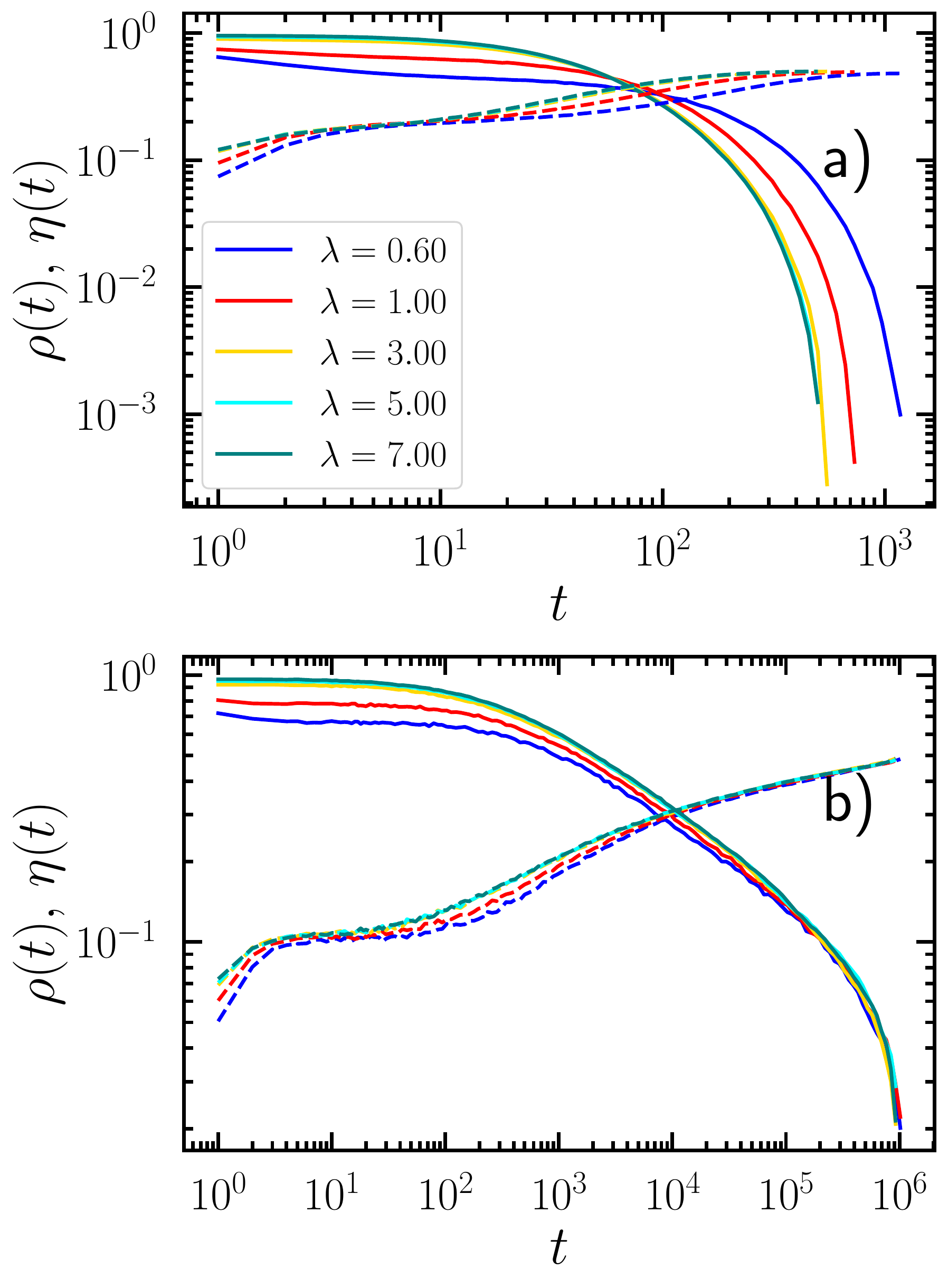}
 \caption{Density of infected nodes $\rho(t)$ (continuous lines) and of
   blocked edges $\eta(t)$ (dashed lines) as a function of time for
   $\gamma=3.2$ and network size $N=10^5$ (a) and $\gamma=2.2$ with
   $N=10^4$ (b).  Since the blocking of an edge may occur in two
   different directions, we consider each edge twice and normalize
   $\eta$ dividing by twice the total number of edges.}
 \label{Confirmation}
\end{figure}

The different decays exhibited depending on $\gamma$ can be traced back
to different topological features of the networks.  It has been shown
that for $\gamma>5/2$ the epidemic transition in SIS is triggered by the
hub (the node with the largest degree $q_\mathrm{max}$) and its direct
neighbors, which singlehandedly can keep alive the dynamics for very
large periods of time, by repeatedly reinfecting each other.  The
elimination of backtracking paths imposed by the NBSIS model forbids
these reinfection events and thus completely destroys this triggering
mechanism.  On the other hand, for $\gamma<5/2$ the transition is
triggered by a different subset of nodes, identified with the $K$-core
of maximum index~\cite{Castellano2012}.  This subgraph is composed of
nodes with a fairly high number of mutual interconnections. It therefore
provides many possible paths through which infection can propagate from
one node to another, circumventing the cancellation of direct
backtracking paths.  Although these alternative paths are not sufficient
to sustain indefinitely the epidemic, they permit the slow decay toward
the absorbing state.

We can obtain a further confirmation of this picture by reinterpreting
the NBSIS model in terms of coupled SIS and dynamic bond percolation
processes~\cite{Callaway2000}. In this sense, when a node $i$ infects
nodes $j$ for the first time, this dynamic step blocks the edge pointing
from $j$ to $i$ for further infections until $i$ is again infected by a
node $m \neq j$.  Accordingly, an infection event from $i$ to $j$ can
unblock an edge from $k\neq j$, previously blocked due to an infection
event at a past time from $k$ to $j$. Thus, the NBSIS dynamics behave as
a SIS model in which edges are blocked and unblocked in a
percolation-like process due to infection events between pairs of
nodes. Notice that this percolation process is intrinsically different
from the standard one~\cite{Callaway2000} since the blocking is dynamic
and correlated. In this framework, it is easy to see that a steady
active state can only be achieved when the density of blocked edges
reaches a steady state of not very large value.  In
Fig.~\ref{Confirmation} we plot the normalized density of blocked edges
$\eta(t)$ and the prevalence $\rho(t)$ as a function of time, for
different values of $\gamma$ and $\lambda$.  We observe that during the
transient state also the density of blocked states is approximately
constant over time. However, the pseudo-steady state eventually ends,
the density of blocked states resumes growth and attains, when the
absorbing state is reached, values close to $\eta=1/2$ which corresponds
to each edge in the network blocked in one direction.  This occurs for
both large and for small values of $\gamma$.

\section{Discussion}
\label{sec:discussion}

Message passing (MP) methods are a powerful tool that can be
successfully applied to a variety of dynamical models on complex
networks. They rely on the neglect of ``echo chambers'' or backtracking
events, in which a set of nodes can influence each other
repeatedly. Dynamical paths are strictly non-backtracking for models
such as percolation or the SIR epidemic spreading model, and therefore
it is natural that a MP approach provides a correct description in such
situations. Such a description provides the thresholds in terms of the
largest eigenvalue of the Hashimoto or non-backtracking matrix.

In other situations, such as the one represented by the SIS model,
backtracking paths represent a crucial ingredient of the dynamics.  They
are at the core of the mechanism that keeps epidemics active in
power-law distributed networks with low level of heterogeneity (i.e. for
a large degree exponent), in which repeated reinfection events between
the hub and its nearest neighbors are able to keep the infection alive
for large intervals of time, and to propagate it to the rest of the
network.  Disallowing such backtracking paths leads inevitably to the
elimination of the steady state of the SIS model. In this case, a better
description is provided by quenched mean-field (QMF) theory, predicting
the threshold to be the inverse of the largest eigenvalue of the
adjacency matrix of the network.

In the present paper we have reconsidered the claims made in
Ref.~\cite{Shrestha2015} concerning the applicability of the MP method
to the description of the SIS dynamics. We have shown that, while no big
difference exists for many real networks, in synthetic uncorrelated ones
the threshold predictions of MP and QMF theories are widely different in
the limit of large networks for mildly heterogeneous networks (for
$\gamma > 5/2$), in agreement with theoretical expectations.  Direct
simulations of the SIS dynamics show that in this same regime, the
numerical threshold of the SIS process is in better agreement with the
QMF prediction than with the MP one, which is not a bound for the true
threshold.

\begin{figure}
  \centering
  \includegraphics[width=\columnwidth]{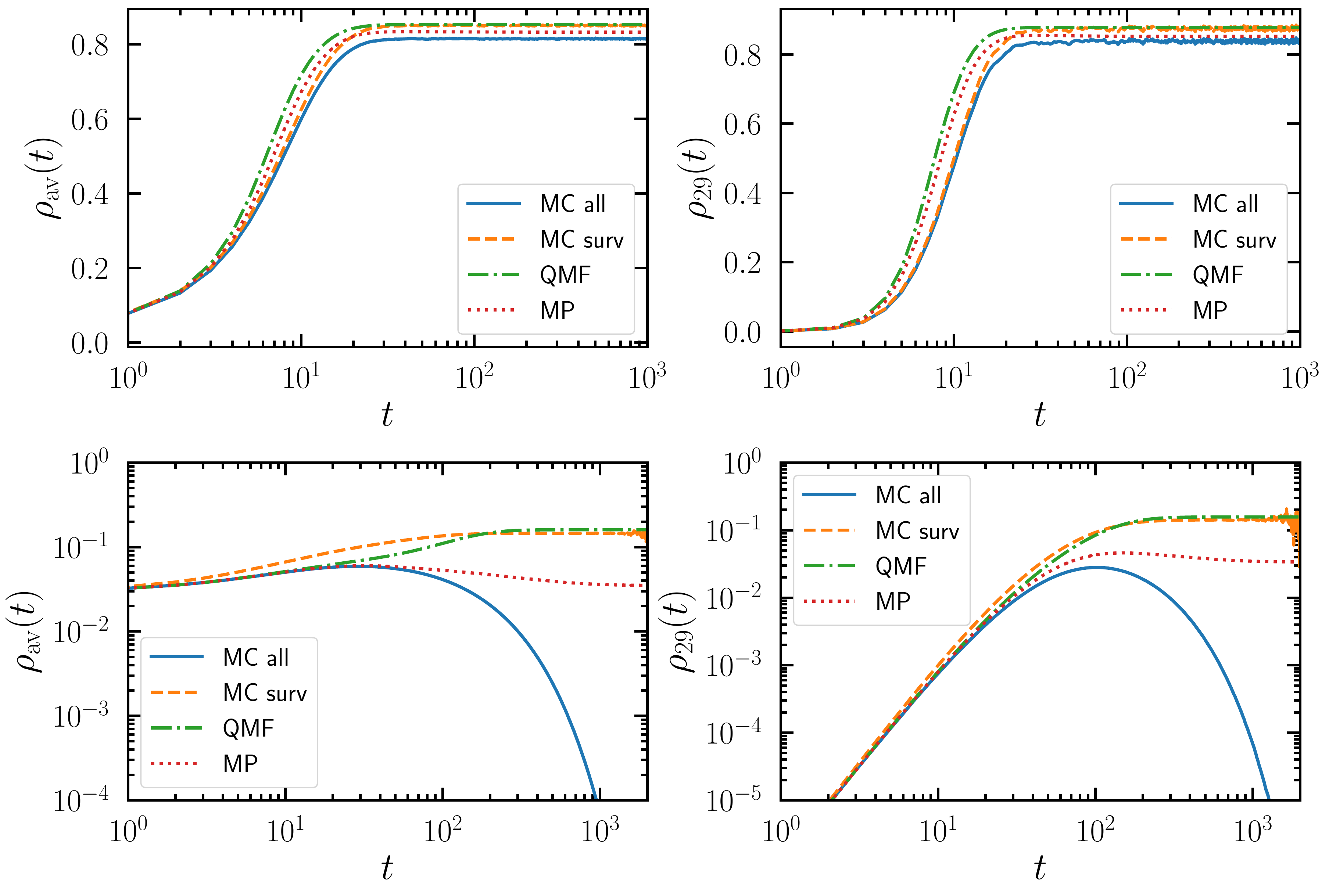}
  \caption{Prevalence as a function of time for a SIS process in the
    Karate club network. Data correspond to the prevalence averaged
    over all nodes (left column) and to the prevalence of node $29$ (right
    column). We compare the results of Monte Carlo simulations,
    performing averages over all runs and over surviving runs, with the
    QMF and MP results obtained from integrating the corresponding
    nonlinear differential equations. SIS parameters are $\mu=0.05$ and
    $\beta=0.1$ (upper row), away from the critical point, and
    $\beta=0.01$ (lower row), close to the critical point. In the initial
    condition the node $0$ is infected and all the others are
    susceptible. Monte Carlo results are averaged over $1000$
    independent runs.}
  \label{ComparisonMCTheories}
\end{figure}

While we have demonstrated that MP does not provide a better estimate of
the SIS threshold than QMF, in Ref.~\cite{Shrestha2015} it is also
argued that MP provides a better approximation to the probability that
individual nodes are infected at a given time. In
Fig.~\ref{ComparisonMCTheories} we clarify this issue by comparing Monte
Carlo estimates of the prevalence of the SIS process as a function of
time on the Karate club network (same as considered in
\cite{Shrestha2015}) with the QMF and MP predictions obtained from
direct numerical integration of the corresponding nonlinear differential
equations. Details are provided in the figure caption.  We observe here
that QMF theory provides a very good approximation to the steady-state
prevalence when this one is measured over \textit{surviving runs}, that
is, over runs that, at given time $t$, have not yet fallen into the
absorbing state. On the other hand, MP theory appears, for systems away
from the critical point, closer to the steady-state prevalence 
computed averaging over all runs, including those already in the absorbing 
state.
Close to the critical point MP yields a steady-state, unobserved in
numerical simulations.  Considering that for all finite networks the
dynamics fall, sooner or later, into the absorbing state, information
about the critical probabilities is to be estimated using only surviving
runs~\cite{Marrobook}.  It is therefore clear that QMF provides a better
estimate of the critical properties of the SIS model. 

In order to better understand the role of backtracking
paths, we have analyzed a SIS-like model in which non-backtracking paths
are suppressed to some extent. Numerical simulations of this model show
that it does not possess a steady state, at odds with SIS behavior,
signaling again that the exclusion of backtracking paths cannot provide
an accurate description of the SIS dynamics.

The consideration of backtracking paths can improve theories for the SIS
dynamics, as they can take into account dynamical correlation
effects~\cite{Boguna09}, see Ref.~\cite{Parshani2010} for an early
attempt.  However, as already stated in~\cite{Karrer2010}, the
development of a message passing method appropriate for SIS still
remains an outstanding open problem.

\begin{acknowledgments}
  We acknowledge financial support from the Spanish MINECO, under project
  FIS2016-76830-C2-1-P.  R.P.-S. acknowledges additional financial support
  from ICREA Academia, funded by the Generalitat de Catalunya.
\end{acknowledgments}

\bibliography{Resubmit_Clean}

\end{document}